\documentclass{article}%
\usepackage{amsmath}
\usepackage{amsfonts}
\usepackage{amssymb}
\usepackage{graphicx}%
\providecommand{\U}[1]{\protect\rule{.1in}{.1in}}

\begin{document}

\title{Entropic Dynamics: Mechanics without Mechanism}
\author{Ariel Caticha\\{\small Physics Department, University at Albany-SUNY, Albany, NY 12222, USA.}}
\date{}
\maketitle

\begin{abstract}
Entropic Dynamics is a framework in which dynamical laws such as those that
arise in physics are derived as an application of entropic methods of
inference. No underlying action principle is postulated. Instead, the dynamics
is driven by entropy subject to constraints reflecting the information that is
relevant to the problem at hand. In this chapter I review the derivation of of
three forms of mechanics. The first is a standard diffusion, the second is a
form of Hamiltonian mechanics, and finally, an argument from information
geometry is used to motivate the particular choice of Hamiltonian that leads
to quantum mechanics.

\end{abstract}

\begin{description}
\item[\textbf{Law without Law:}] \emph{\textquotedblleft The only thing harder
to understand than a law of statistical origin would be a law that is not of
statistical origin, for then there would be no way for it --- or its
progenitor principles --- to come into being.\textquotedblright\ }

\item[\textbf{Two tests:}] \emph{\textquotedblleft No test of these views
looks like being someday doable, nor more interesting and more instructive,
than a derivation of the structure of quantum theory... No prediction lends
itself to a more critical test than this, that every law of physics, pushed to
the extreme, will be found statistical and approximate, not mathematically
perfect and precise.\textquotedblright\ }

\item \hspace*{\fill}\emph{J. A. Wheeler }\cite{Wheeler Zurek 1983}
\end{description}

\section{Introduction}

The drive to explain nature has always led us to seek the mechanisms hidden
behind the phenomena. Descartes, for example, claimed to explain the motion of
planets as being swept along in the flow of some vortices. The model did not
work very well but at least it gave the illusion of a mechanical explanation
and thereby satisfied a deep psychological need. Newton's theory fared much
better. He took the important step of postulating that gravity was a universal
force acting at a distance but he abstained from offering any mechanical
explanation --- a stroke of genius immortalized in his famous
\textquotedblleft\emph{hypotheses non fingo}.\textquotedblright\ At first
there were objections. Huygens, for instance, recognized the undeniable value
of Newton's achievement but was nevertheless deeply disappointed: the theory
\emph{works} but it does not \emph{explain.} And Newton agreed: any action at
a distance would represent \textquotedblleft\emph{so great an absurdity...
that no man who has in philosophical matters a competent faculty of thinking
can ever fall into it.}\textquotedblright\ \cite{Newton 1693}

Over the following 18th century, however, impressed by the many successes of
Newtonian mechanics, people started to downplay and then even forget their
qualms about the absurdity\ of action at a distance. Mechanical explanations
were, of course, still required but the very meaning of what counted as
\textquotedblleft mechanical\textquotedblright\ suffered a gradual but
irreversible shift. It no longer meant \textquotedblleft caused by contact
forces\textquotedblright\ but rather \textquotedblleft described according to
Newton's laws.\textquotedblright\ Over time Newtonian forces, including those
mysterious actions at a distance, became \textquotedblleft
real\textquotedblright\ which qualified them to count as the causes behind phenomena.

But this did not last too long. With Lagrange, Hamilton, and Maxwell, the
notion of force started to lose some of its fundamental status. Indeed, after
Maxwell succeeded in extending the principles of dynamics to include the
electromagnetic field, the meaning of `mechanical explanation' changed once
again. It no longer meant identifying the Newtonian forces, but rather
identifying the right Lagrangian, the right action principle. Thus, today
gravity is no longer explained through a force but through the curvature of
space-time. And the concept of force finds no place in quantum mechanics where
interactions are described as the evolution of abstract vectors in Hilbert space.

Our goal in this chapter is to derive useful dynamical models without invoking
underlying mechanisms. This does not mean that such mechanisms do not exist,
for all we know they might. It is just that useful models can be constructed
without having to go through the trouble of keeping track of a myriad of
microscopic details that often turn out to be ultimately irrelevant. The idea
can be illustrated by contrasting the two very different ways in which the
theory of Brownian motion was originally derived by Smoluchowski and by
Einstein. In Smoluchowski's approach one keeps track of the microscopic
details of molecular collisions through a stochastic Langevin equation and a
macroscopic effective theory is then derived by taking suitable averages. In
Einstein's approach, on the other hand, one focuses directly on those pieces
of information that turn out to be relevant for the prediction of macroscopic
effects. The advantage of Einstein's approach is the simplicity that arises
from not having to keep track of irrelevant details that are eventually washed
out when taking the averages. The challenge, of course, is to identify those
pieces of information that happen to be relevant.

Our argument proceeds by stages. We will exhibit three explicit examples.
Starting with the simplest, we first tackle a standard diffusion, which allows
us to introduce and discuss the non-trivial concept of time. In the next stage
we discuss the derivation of Hamiltonian dynamics as a non-dissipative type of
diffusion. Finally, in a further elaboration invoking notions of information
geometry, we identify the particular Hamiltonian that leads to quantum mechanics.

Ever since its origin in 1925 the interpretation of quantum mechanics has been
a notoriously controversial subject. The central question revolves around the
interpretation of the quantum state, the wave function: does it represent the
actual real state of the system (its \emph{ontic} state) or does it represent
a state of knowledge about the system (an \emph{epistemic} state)? Excellent
reviews with extended references to the literature are given in \cite{Stapp
1972}-\cite{Leifer 2014}. However, the urge to seek underlying mechanisms has
motivated a huge effort towards identifying hidden variables and sub-quantum
levels of reality from which an effective quantum dynamics might emerge. (See
\emph{e.g.}, \cite{Nelson 1985}-\cite{EmQm 2015}.) In this work we adopt the
point of view that quantum theory itself provides us with the most important
clue: the goal of quantum mechanics is to calculate probabilities.
\emph{Therefore}, by its very nature, quantum mechanics must be an instance of
Bayesian and entropic inference.

Entropic Dynamics (ED) provides a framework for deriving dynamical laws as an
application of entropic methods \cite{Caticha 2010}-\cite{Caticha 2015}. The
use of entropy as a tool for inference can be traced to E. T. Jaynes
\cite{Jaynes 1957}-\cite{Jaynes 2003}. For a pedagogical overview and further
references see \cite{Caticha 2012}.

The literature on the attempts to reconstruct quantum mechanics is vast and
there are several approaches that like ED\ are also based on information
theory (see \emph{e.g.}, \cite{Wootters 1981}-\cite{DAriano 2017}). Since all
these approaches must sooner or later converge to the same Schr\"{o}dinger
equation it is inevitable that they will show similarities but there are,
however, important differences. What distinguishes ED is a strict adherence to
Bayesian and entropic methods without invoking mechanisms operating at some
deeper sub-quantum level.

This chapter is basically a self-contained review of material developed in
\cite{Caticha 2010}-\cite{Caticha 2015} but with one significant change. The
gist of Entropic Dynamics is that the system of interest (\emph{e.g.},
particles or fields) undergoes a diffusion with a special systematic drift
which is ultimately responsible for such quintessential quantum phenomena as
interference and entanglement. A central idea introduced in \cite{Caticha
2010}, but not fully developed in later publications, is that this peculiar
drift is itself of entropic origin. More precisely, the drift follows the
gradient of the entropy of some extra variables which remain hidden or at
least inaccessible. Here we intend to pursue this fully entropic
interpretation of the drift while taking advantage of later insights that
exploit ideas from information geometry.

An important feature of ED is a central concern with the nature of time. In ED
\textquotedblleft entropic\textquotedblright\ time is \emph{designed} to keep
track of change. The construction of entropic time involves several
ingredients. One must define the notion of `instants'; show that these
instants happen to be ordered; and finally one must define a convenient
measure of the duration or interval between the successive instants. As one
might have expected, in an entropic approach to time it turns out that an
arrow of time is generated automatically. Entropic time is intrinsically directional.

Here we will focus on the derivation of the Schr\"{o}dinger equation but the
ED approach has been applied to a variety of other topics in quantum mechanics
that will not be reviewed here. These include the quantum measurement problem
\cite{Johnson Caticha 2011}\cite{Vanslette Caticha 2016}; momentum and
uncertainty relations \cite{Nawaz Caticha 2011}; the Bohmian limit
\cite{Bartolomeo Caticha 2015}\cite{Bartolomeo Caticha 2016} and the classical
limit \cite{Demme Caticha 2016}; the extensions to curved spaces \cite{Nawaz
et al 2015} and to relativistic fields \cite{Ipek Caticha 2014}\cite{Ipek
Abedi Caticha 2016}; and the derivation of symplectic and complex structures
from information geometry \cite{Caticha 2017}.

\section{The statistical model}

As in any inference problem we must first decide on the subject matter ---
what are we talking about? We consider $N$ particles living in a flat
Euclidean space $\mathbf{X}$ with metric $\delta_{ab}$. Here is our first
assumption: \emph{The particles have definite positions }$x_{n}^{a}$\emph{,
collectively denoted by }$x$\emph{, and it is their unknown values that we
wish to infer.} (The index $n$ $=1\ldots N$ labels the particles and $a=1,2,3$
its spatial coordinates.) For $N$ particles the configuration space is
$\mathbf{X}_{N}=\mathbf{X}\times\ldots\times\mathbf{X}$. The fact that
positions are unknown implies that the probability distribution $\rho(x)$ will
be the focus of our attention.

Incidentally, this already addresses that old question of determinism vs.
indeterminism that has troubled quantum mechanics from the outset. In the
standard view quantum theory is considered to be an extension of classical
mechanics and therefore deviations from causality demand an explanation. In
contrast, according to the entropic view, quantum mechanics is an example of
entropic inference, a framework designed to handle insufficient information.
From this entropic perspective indeterminism requires no explanation;
uncertainty and probabilities are the expected norm. It is certainty and
determinism that demand explanations \cite{Demme Caticha 2016}.

The assumption that the particles have definite positions that happen to be
unknown is a not an innocent statement. It represents a major shift away from
the standard Copenhagen interpretation. Such a departure is not in itself a
problem: our goal here is not to justify any particular interpretation of
quantum mechanics but to reproduce its empirical success. According to the
Copenhagen interpretation observable quantities such as positions, do not in
general have definite values; they can become definite but only as the
immediate result of a measurement. ED differs in that positions play a very
special role: at all times particles have definite positions which define the
ontic state of the system. The wave function, on the other hand, is a purely
epistemic notion; it defines our epistemic state about the system. In ED ---
just as in the Copenhagen interpretation --- other observables such as energy
or momentum do not in general have definite values; their values are created
by the act of measurement. These other quantities are epistemic in that they
do not reflect properties of the particles but of the wave function. Thus
positions are ontic while energies and momenta are epistemic. This point
deserves to be emphasized: in the ED description of the double slit
experiment, we might not know which slit the particle goes through, but it
definitely goes through one or the other.

The second assumption is that \emph{in addition to the particles there also
exist some other variables denoted }$y$. The assumption does not appear
extreme.\footnote{As we shall later discuss in more detail in the concluding
remarks, the $y$ variables, although hidden\ from our view, do not at all play
the technical role usually ascribed to \textquotedblleft hidden
variables\textquotedblright.} First, the world definitely contains more stuff
than the $N$ particles we happen to be currently interested in and, second, it
turns out that our main conclusions are very robust in that they turn out to
be largely independent of most particular details about these $y$ variables.
We only need to assume that the $y$ variables are themselves uncertain and
that their uncertainty, described by some probability distribution $p(y|x)$,
depends on the location $x$ of the particles. As we shall soon see it is the
entropy of the distribution $p(y|x)$ that plays a significant role in defining
the dynamics of $x$. Other details of the distribution $p(y|x)$ turn out to be
irrelevant. For later reference, the entropy $S(x)$ of $p(y|x)$ relative to an
underlying measure $q(y)$ on the space of $y$ variables is
\begin{equation}
S(x)=-\int dy\,p(y|x)\log\frac{p(y|x)}{q(y)}~. \label{entropy a}%
\end{equation}
For notational simplicity in multidimensional integrals such as
(\ref{entropy a}) we will write $dy$ instead of $d^{n}y$. Note also that
$S(x)$ is a scalar function on the configuration space $\mathbf{X}_{N}$.

Once the microstates $(x,y)\in\mathbf{X}_{N}\times\mathbf{Y}$ are identified
we proceed to tackle the dynamics. Our goal is to find the probability density
$P(x^{\prime}|x)$ for a change from an initial position $x\in\mathbf{X}_{N}$
to a new neighboring $x^{\prime}\in\mathbf{X}_{N}$. Since both $x^{\prime}$
and the associated $y^{\prime}$ are unknown the relevant space is not just
$\mathbf{X}_{N}$ but $\mathbf{X}_{N}\times\mathbf{Y}$ and, therefore, the
distribution we seek is the joint distribution $P(x^{\prime},y^{\prime}|x,y)$.
This is found by maximizing the appropriate entropy,
\begin{equation}
\mathcal{S}[P,Q]=-\int dx^{\prime}dy^{\prime}\,P(x^{\prime},y^{\prime
}|x,y)\log\frac{P(x^{\prime},y^{\prime}|x,y)}{Q(x^{\prime},y^{\prime}|x,y)}~,
\label{entropy joint a}%
\end{equation}
relative to the joint prior $Q(x^{\prime},y^{\prime}|x,y)$ and subject to the
appropriate constraints.

\subsubsection*{Choosing the prior}

In eq.(\ref{entropy joint a}) the prior $Q(x^{\prime},y^{\prime}|x,y)$
expresses our beliefs --- or more precisely, the beliefs of an ideally
rational agent --- \emph{before} any information about the specific motion is
taken into account. We adopt a prior that represents a state of considerable
ignorance: knowledge of $x$s tells us nothing about $y$s and vice versa. This
is represented by a product,
\begin{equation}
Q(x^{\prime},y^{\prime}|x,y)=Q(x^{\prime}|x)Q(y^{\prime}|y)~.
\end{equation}
The prior $Q(y^{\prime}|y)$ for the $y$ variables is chosen to be a uniform
distribution. If the volume element in the space $\mathbf{Y}$ is proportional
to $q(y)dy$ then $Q(y^{\prime}|y)\propto q(y^{\prime})$. The measure $q(y)$
need not be specified further. The prior $Q(x^{\prime}|x)$ for the $x$
variables is considerably more informative.

In a \textquotedblleft mechanics without a mechanism\textquotedblright\ one
does not explain why motion happens. The goal instead is to produce an
estimate of what kind of motion one might reasonably expect. \emph{The central
piece of dynamical information is that the particles follow trajectories that
are continuous.} This represents yet another significant deviation from the
early historical development of quantum mechanics which stressed discreteness
and discontinuity. Fortunately, in modern versions of quantum mechanics these
two aspects are de-emphasized. Ever since Schr\"{o}dinger's pioneering work
the discreteness of quantum numbers is not mysterious --- certainly not more
mysterious than the discrete resonances of classical vibrations. And the
discontinuities implicit in abrupt quantum jumps, a relic from Bohr's old
quantum theory, just do not exist; they were eventually recognized as
unnecessary and effectively discarded as soon as Schr\"{o}dinger formulated
his equation. Incidentally, there is another source of discontinuity --- the
so-called wave function collapse. Its mystery disappears the moment one
recognizes the epistemic nature of the wave function: the collapse is not
physical but merely an instance of probability updating \cite{Johnson Caticha
2011}\cite{Vanslette Caticha 2016}.

The assumption of continuity represents a tremendous simplification because it
implies that motion can be analyzed as the accumulation of many
infinitesimally short steps. We adopt a prior $Q(x^{\prime}|x)$ that
incorporates the information that the particles take steps that are
infinitesimally short, that reflects translational and rotational invariance,
but is otherwise maximally ignorant about any correlations. Such a prior can
itself be derived from the principle of maximum entropy. Indeed, maximize
\begin{equation}
S[Q]=-\int dx^{\prime}\,Q(\Delta x)\log\frac{Q(\Delta x)}{\mu}~,
\end{equation}
where $\Delta x_{n}^{a}=x_{n}^{\prime a}-x_{n}^{a}$, relative to the uniform
measure $\mu=\,$const., subject to normalization, and $N$ independent
constraints that enforce short steps and rotational invariance,
\begin{equation}
\langle\delta_{ab}\Delta x_{n}^{a}\Delta x_{n}^{b}\rangle=\kappa_{n}%
~,\quad(n=1\ldots N)~,
\end{equation}
where $\kappa_{n}$ are small constants. The result is a product of Gaussians,
\begin{equation}
Q(x^{\prime}|x)\propto\exp-\frac{1}{2}%
{\displaystyle\sum\nolimits_{n}}
\alpha_{n}\delta_{ab}\Delta x_{n}^{a}\Delta x_{n}^{b}~.
\end{equation}
The Lagrange multipliers $\alpha_{n}$ are constants that may depend on the
index $n$ in order to describe non-identical particles. They will eventually
be taken to infinity in order to enforce infinitesimally short steps.

The choice of a Gaussian prior turns out to be natural, not just because it
arises in an informational context, but also because, as described by the
central limit theorem as in the theory of errors, it arises whenever a
\textquotedblleft macroscopic\textquotedblright\ effect is the result of the
accumulation of a large number of \textquotedblleft
microscopic\textquotedblright\ contributions.

Since proportionality constants have no effect on the entropy maximization,
our choice for the joint prior is
\begin{equation}
Q(x^{\prime},y^{\prime}|x,y)=q(y^{\prime})\exp-\frac{1}{2}%
{\displaystyle\sum\nolimits_{n}}
\alpha_{n}\delta_{ab}\Delta x_{n}^{a}\Delta x_{n}^{b}~. \label{prior}%
\end{equation}
Now we are ready to specify the constraints.

\subsubsection*{The constraints}

We first write the posterior as a product,%
\begin{equation}
P(x^{\prime},y^{\prime}|x,y)=P(x^{\prime}|x,y)P(y^{\prime}|x^{\prime},x,y)\,,
\label{joint prob}%
\end{equation}
and we specify the two factors separately. We require that the new $x^{\prime
}$ depends only on $x$, so we set $P(x^{\prime}|x,y)=P(x^{\prime}|x)$ which is
the transition probability we wish to find.\ The new $x^{\prime}$ is
independent of the actual values of $y$ or $y^{\prime}$ but, as we shall soon
see, it does depend on their entropies.

As for the second factor in (\ref{joint prob}), we require that $P(y^{\prime
}|x^{\prime},x,y)=p(y^{\prime}|x^{\prime})$ --- the uncertainty in $y^{\prime
}$ depends only on $x^{\prime}$. Therefore, the first constraint is that the
joint posterior be of the form%
\begin{equation}
P(x^{\prime},y^{\prime}|x)=P(x^{\prime}|x)p(y^{\prime}|x^{\prime})~.
\label{constraint p}%
\end{equation}
To implement this constraint substitute (\ref{constraint p}) into the joint
entropy (\ref{entropy joint a}) to get
\begin{equation}
\mathcal{S}[P,Q]=-\int dx^{\prime}\,P(x^{\prime}|x)\log\frac{P(x^{\prime}%
|x)}{Q(x^{\prime}|x)}+\int dx^{\prime}\,P(x^{\prime}|x)S(x^{\prime})~,
\label{entropy joint b}%
\end{equation}
where $S(x)$ is given in eq.(\ref{entropy a}).

Having specified the prior and the constraints the ME method takes over. We
vary $P(x^{\prime}|x)$ to maximize (\ref{entropy joint b})\ subject to
normalization. The result is
\begin{equation}
P(x^{\prime}|x)=\frac{1}{\zeta}\exp\left[  S(x^{\prime})-\frac{1}{2}%
{\displaystyle\sum\nolimits_{n}}
\alpha_{n}\delta_{ab}\Delta x_{n}^{a}\Delta x_{n}^{b}\right]  ~,
\label{Prob x'/x}%
\end{equation}
where $\zeta$ is a normalization constant. In eq.(\ref{Prob x'/x}) it is clear
that infinitesimally short steps are obtained in the limit $\alpha
_{n}\rightarrow\infty$. It is therefore useful to Taylor expand,
\begin{equation}
S(x^{\prime})=S(x)+\sum\nolimits_{n}\Delta x_{n}^{a}\frac{\partial S}{\partial
x_{n}^{a}}+\ldots
\end{equation}
and rewrite $P(x^{\prime}|x)$ as
\begin{equation}
P(x^{\prime}|x)=\frac{1}{Z}\exp\left[  -\frac{1}{2}%
{\displaystyle\sum\nolimits_{n}}
\alpha_{n}\,\delta_{ab}\left(  \Delta x_{n}^{a}-\langle\Delta x_{n}^{a}%
\rangle\right)  \left(  \Delta x_{n}^{b}-\langle\Delta x_{n}^{b}%
\rangle\right)  \right]  ~, \label{Prob x'/x b}%
\end{equation}
where $Z$ is the new normalization constant. A generic displacement $\Delta
x_{n}^{a}=x_{n}^{\prime a}-x_{n}^{a}$ can be expressed as an expected drift
plus a fluctuation,
\begin{equation}
\Delta x_{n}^{a}=\left\langle \Delta x_{n}^{a}\right\rangle +\Delta w_{n}%
^{a}\,\,,\quad\text{where\quad}\langle\Delta x_{n}^{a}\rangle=\frac{1}%
{\alpha_{n}}\delta^{ab}\frac{\partial S}{\partial x_{n}^{b}}~,
\label{ED drift}%
\end{equation}%
\begin{equation}
\left\langle \Delta w_{n}^{a}\right\rangle =0\quad\text{and}\quad\langle\Delta
w_{n}^{a}\Delta w_{n}^{b}\rangle=\frac{1}{\alpha_{n}}\delta^{ab}~.
\label{ED fluctuations}%
\end{equation}
These equations show that as $\alpha_{n}\rightarrow\infty$, for very short
steps, the dynamics is dominated by fluctuations $\Delta w_{n}^{a}$ which are
of order $O(\alpha_{n}^{-1/2})$, while the drift $\langle\Delta x_{n}%
^{a}\rangle$ is much smaller, only of order $O(\alpha_{n}^{-1})$. Thus, just
as in Brownian motion, the trajectory is continuous but not differentiable. In
ED particles have definite positions but their velocities are completely
undefined. Notice also that the particle fluctuations are isotropic and
independent of each other. The directionality of the motion and correlations
among the particles are introduced by a systematic drift along the gradient of
the entropy of the $y$ variables.\ 

The introduction of the auxiliary $y$ variables, which has played a central
role of conveying relevant information about the drift, deserves a comment. It
is important to realize that the same information can be conveyed through
other means, which demonstrates that quantum mechanics, as an effective
theory, turns out to be fairly robust\footnote{In the case of physics, the $y$
variables might for example describe the microscopic structure of spacetime.
In the case of an economy, they may describe the hidden neural processes of
various agents. It is a strength of the ED formulation that details about
these auxilliary variables are not needed.} \cite{Bartolomeo Caticha
2015}\cite{Bartolomeo Caticha 2016}. It is possible, for example, to avoid the
$y$ variables altogether and invoke a drift potential directly (see
\emph{e.g.}, \cite{Caticha 2014}-\cite{Caticha 2015}). The advantage of an
explicit reference to some vaguely defined $y$ variables is their mere
existence may be sufficient to account for drift effects.

\section{Entropic time}

Having obtained a prediction, given by eq.(\ref{Prob x'/x b}), for what motion
to expect in one infinitesimally short step we now consider motion over finite
distances. This requires the introduction of a parameter $t$, to be called
\emph{time}, as a book-keeping tool to keep track of the accumulation of
successive short steps. Since the rules of inference are silent on the subject
of time we need to justify why the parameter $t$ deserves to be called time.\ 

The construction of time involves three ingredients. First, we must identify
something that one might call an \textquotedblleft instant\textquotedblright;
second, it must be shown that these instants are ordered; and finally, one
must introduce a measure of separation between these successive instants ---
one must define \textquotedblleft duration.\textquotedblright\ Since the
foundation for any theory of time is the theory of change --- that is,
dynamics --- the notion of time constructed below will reflect the inferential
nature of entropic dynamics. Such a construction we will call \emph{entropic
time}.

\subsection{Time as an ordered sequence of instants}

Entropic dynamics is given by a succession of short steps described by
$P(x^{\prime}|x)$, eq.(\ref{Prob x'/x b}). Consider, for example, the $i$th
step which takes the system from $x=x_{i-1}$ to $x^{\prime}=x_{i}$.
Integrating the joint probability, $P(x_{i},x_{i-1})$, over $x_{i-1}$ gives
\begin{equation}
P(x_{i})=\int dx_{i-1}P(x_{i},x_{i-1})=\int dx_{i-1}P(x_{i}|x_{i-1}%
)P(x_{i-1})~. \label{CK a}%
\end{equation}
This equation follows directly from the laws of probability, it involves no
assumptions and, therefore, it is sort of empty. To make it useful something
else must be added. Suppose we interpret $P(x_{i-1})$ as the probability of
different values of $x_{i-1}$ \emph{at one \textquotedblleft
instant\textquotedblright\ labelled }$t$, then we can interpret $P(x_{i})$ as
the probability of values of $x_{i}$ \emph{at the next \textquotedblleft
instant\textquotedblright} which we will label $t^{\prime}$. Writing
$P(x_{i-1})=\rho(x,t)$ and $P(x_{i})=\rho(x^{\prime},t^{\prime})$ we have
\begin{equation}
\rho(x^{\prime},t^{\prime})=\int dx\,P(x^{\prime}|x)\rho(x,t)~. \label{CK b}%
\end{equation}
Nothing in the laws of probability leading to eq.(\ref{CK a}) forces the
interpretation (\ref{CK b}) on us --- this is the additional ingredient that
allows us to construct time and dynamics in our model. Equation (\ref{CK b})
defines the notion of \textquotedblleft instant\textquotedblright:\ If the
distribution $\rho(x,t)$\ refers to one instant $t$, then the distribution
$\rho(x^{\prime},t^{\prime})$\ generated by $P(x^{\prime}|x)$ defines what we
mean by the \textquotedblleft next\textquotedblright\ instant $t^{\prime}$.
Iterating this process defines the dynamics. Entropic time is constructed
instant by instant: $\rho(t^{\prime})$ is constructed from $\rho(t)$,
$\rho(t^{\prime\prime})$ is constructed from $\rho(t^{\prime})$, and so on.

The construction is intimately related to information and inference. An
`instant' is an informational state that is complete in the sense that it is
specified by the information --- codified into the distributions $\rho(x,t)$
and $P(x^{\prime}|x)$ --- that is sufficient for predicting the next instant.
To put it briefly: \emph{the instant we call the present is such that the
future,\ given the present, is independent of the past}.

In the ED framework the notions of instant and of simultaneity are intimately
related to the distribution $\rho(x)$. It is instructive to discuss this
further. When we consider a single particle at a position $\vec{x}%
=(x^{1},x^{2},x^{3})$ in $3$-d space it is implicit in the notation that the
three coordinates $x^{1}$, $x^{2}$, and $x^{3}$ occur simultaneously --- no
surprises here. Things get a bit more interesting when we describe a system of
$N$ particles by a single point $x=(\vec{x}_{1},\vec{x}_{2},\ldots\vec{x}%
_{N})$ in $3N$-dimensional configuration space. Here it is also implicitly
assumed that all the $3N$ coordinate values refer to the same instant; we take
them to be simultaneous. But this is no longer so trivial because the
particles are located at different places --- we do not mean particle 1 is at
$\vec{x}_{1}$ now, and particle 2 is at $\vec{x}_{2}$ later. There is an
implicit assumption linking the very idea of a configuration space with that
of simultaneity and something similar occurs when we introduce probabilities.
Whether we talk about one particle or about $N$ particles, a distribution such
as $\rho(x)$ describes our uncertainty about the possible configurations $x$
of the system at a given instant. The different values of $x$ refer to the
same instant; they are meant to be simultaneous. And therefore, in ED,\ a
probability distribution $\rho(x)$ provides a criterion of simultaneity.

In a relativistic theory there is a greater freedom in the choice of instants
and this translates into a greater flexibility with the notion of
simultaneity. Conversely, as we have shown elsewhere, the requirement that
these different notions of simultaneity be consistent with each other places
strict constraints on the allowed forms of ED \cite{Ipek Abedi Caticha 2016}.

It is common to use equations such as (\ref{CK b}) to define a special kind of
dynamics, called Markovian, that unfolds in a time defined by some external
clocks. In such a Markovian dynamics the specification of the state at one
instant is sufficient to determine its evolution into the future.\footnote{The
application of entropic dynamics to continuous motion is, of course, heavily
motivated by the application to physics; for all we know space and time form a
continuum. The derivation of discrete Markov models as a form of entropic
dynamics is a subject for future research. But even in the case of an
essentially discrete dynamics (such as e.g. the daily updates of the stock
market) the idealization of continuous evolution is still useful as an
approximation over somewhat longer time scales.} It is important to recognize
that in ED we are not making a Markovian assumption. Although eq.(\ref{CK b})
is formally identical to the Chapman-Kolmogorov equation it is used for a very
different purpose. We do not use (\ref{CK b}) to define a (Markovian) dynamics
in a pre-existing background time because in ED there are no external clocks.
The system is its own clock and (\ref{CK b}) is used both to define the
dynamics and to construct time itself.

\subsection{The arrow of entropic time}

The notion of time constructed according to eq.(\ref{CK b}) is intrinsically
directional. There is an absolute sense in which $\rho(x,t)$\ is prior and
$\rho(x^{\prime},t^{\prime})$\ is posterior. If we wanted to construct a
time-reversed evolution we would write
\begin{equation}
\rho(x,t)=%
{\textstyle\int}
dx^{\prime}\,P(x|x^{\prime})\rho(x^{\prime},t^{\prime})\,,
\end{equation}
where according to the rules of probability theory $P(x|x^{\prime})$ is
related to $P(x^{\prime}|x)$ in eq.(\ref{Prob x'/x b}) by Bayes' theorem,
\begin{equation}
P(x|x^{\prime})=\frac{\rho(x,t)}{\rho(x^{\prime},t^{\prime})}P(x^{\prime}|x)~.
\label{Bayes thm}%
\end{equation}
This is not, however, a mere exchange of primed and unprimed quantities. The
distribution $P(x^{\prime}|x)$, eq.(\ref{Prob x'/x b}), is a Gaussian derived
from the maximum entropy method. In contrast, the time-reversed $P(x|x^{\prime
})$ is given by Bayes' theorem, eq.(\ref{Bayes thm}), and is not in general
Gaussian. The asymmetry between the inferential past and the inferential
future is traced to the asymmetry between priors and posteriors.

The puzzle of the arrow of time (see \emph{e.g.} \cite{Price 1996}\cite{Zeh
2002}) has been how to explain the asymmetric arrow from underlying symmetric
laws. The solution offered by ED is that there are no underlying laws whether
symmetric or not. The time asymmetry is the inevitable consequence of entropic
inference. From the point of view of ED the challenge is not to explain the
arrow of time but the reverse: how to explain the emergence of symmetric laws
within an entropic framework that is intrinsically asymmetric. As we shall see
below some laws of physics derived from ED, such as the Schr\"{o}dinger
equation, are indeed time-reversible even though entropic time itself only
f{}lows forward.

\subsection{Duration: a convenient time scale}

To complete the construction of entropic time we need to specify the interval
$\Delta t$ between successive instants. The basic criterion is convenience:
\emph{duration is defined so that motion looks simple.} We saw in
eqs.(\ref{ED drift}) and (\ref{ED fluctuations}) that for short steps (large
$\alpha_{n}$) the motion is largely dominated by f{}luctuations. Therefore
specifying $\Delta t$ amounts to specifying the multipliers $\alpha_{n}$ in
terms of $\Delta t$.

The description of motion is simplest when it reflects the symmetry of
translations in space and time. In a flat spacetime this leads us to an
entropic time that resembles Newtonian time in that it flows \textquotedblleft
equably\ everywhere and everywhen.\textquotedblright\ Thus, we choose
$\alpha_{n}$ to be independent of $x$ and $t$, and we choose $\Delta t$ so
that $\alpha_{n}\propto1/\Delta t$. Furthermore, it is convenient to express
the proportionality constants in terms of some particle-specific constants
$m_{n}$ and an overall constant $\eta$ that fixes the units of the $m_{n}$s
relative to the units of time. The result is
\begin{equation}
\alpha_{n}=\frac{m_{n}}{\eta}\frac{1}{\Delta t}~. \label{alpha n}%
\end{equation}
The constants $m_{n}$ will eventually be identified with the particle masses
and the constant $\eta$ will be regraduated into $\hbar$.

\section{The information metric of configuration space}

Before we proceed to study the dynamics defined eq.(\ref{Prob x'/x b}) with
its corresponding notion of entropic time it is useful to consider the
geometry of the $N$-particle configuration space, $\mathbf{X}_{N}$. We have
assumed that the geometry of the single particle spaces $\mathbf{X}$ is
described by the Euclidean metric $\delta_{ab}$. We can expect that the
$N$-particle configuration space, $\mathbf{X}_{N}=\mathbf{X}\times\ldots
\times\mathbf{X}$, will also be flat, but a question remains about the
relative scales associated to each $\mathbf{X}$ factor. Information geometry
provides the answer.

To each point $x\in\mathbf{X}_{N}$ there corresponds a probability
distribution $P(x^{\prime}|x)$. This means that $\mathbf{X}_{N}$ is a
statistical manifold the geometry of which is uniquely determined (up to an
overall scale factor) by the information metric,%
\begin{equation}
\gamma_{AB}=C\int dx^{\prime}\,P(x^{\prime}|x)\frac{\partial\log P(x^{\prime
}|x)}{\partial x^{A}}\frac{\partial\log P(x^{\prime}|x)}{\partial x^{B}}~.
\label{gamma C}%
\end{equation}
Here the upper case indices label both the particle and its coordinate,
$x^{A}=x_{n}^{a}$, and $C$ is an arbitrary positive constant (see \emph{e.g.},
\cite{Caticha 2012}\cite{Amari 1985}). Substituting eqs.(\ref{Prob x'/x b})
and (\ref{alpha n}) into (\ref{gamma C}) in the limit of short steps
($\alpha_{n}\rightarrow\infty$) yields
\begin{equation}
\gamma_{AB}=\frac{Cm_{n}}{\eta\Delta t}\delta_{nn^{\prime}}\,\delta_{ab}%
=\frac{Cm_{n}}{\eta\Delta t}\delta_{AB}~. \label{gamma AB}%
\end{equation}
Note the divergence as $\Delta t\rightarrow0$. Indeed, as $\Delta
t\rightarrow0$ the distributions $P(x^{\prime}|x)$ and $P(x^{\prime}|x+\Delta
x)$ become more sharply peaked and they become increasingly easier to
distinguish. This leads to an increasing information distance, $\gamma
_{AB}\rightarrow\infty$. To define a distance that remains useful for
arbitrarily small $\Delta t$ we choose $C\propto\Delta t$. In fact, since
$\gamma_{AB}$ will always appear in the combination $\gamma_{AB}\Delta t/C$,
it is best to define the \textquotedblleft mass\textquotedblright\ tensor,
\begin{equation}
m_{AB}=\frac{\eta\Delta t}{C}\gamma_{AB}=m_{n}\delta_{AB}~. \label{mass a}%
\end{equation}
and its inverse,
\begin{equation}
m^{AB}=\frac{C}{\eta\Delta t}\gamma^{AB}=\frac{1}{m_{n}}\delta^{AB}~.
\label{mass b}%
\end{equation}

Thus, up to overall constants the metric of configuration space is the mass
tensor. This result may be surprising. Ever since the work of Heinrich Hertz
in 1894 \cite{Lanczos 1970} it has been standard practice to describe the
motion of systems with many particles as the motion of a single point in an
abstract space --- the configuration space. Choosing the geometry of this
configuration space had so far been regarded as a matter of convenience and
was suggested through an examination of the kinetic energy of the system. In
contrast, in ED there is no choice. Up to a global scale the metric follows
uniquely determined by information geometry.

To recap our results so far: with the multipliers $\alpha_{n}$ chosen
according to (\ref{alpha n}), the dynamics given by $P(x^{\prime}|x)$
in\ (\ref{Prob x'/x b}) is a standard Wiener process. A generic displacement,
eq.(\ref{ED drift}), is
\begin{equation}
\Delta x^{A}=b^{A}\Delta t+\Delta w^{A}~, \label{Delta x}%
\end{equation}
where $b^{A}(x)$ is the drift velocity,
\begin{equation}
\langle\Delta x^{A}\rangle=b^{A}\Delta t\quad\text{with}\quad b^{A}=\eta
m^{AB}\partial_{B}S~, \label{drift velocity}%
\end{equation}
and the uncertainty $\Delta w^{A}$ is given by
\begin{equation}
\langle\Delta w^{A}\rangle=0\quad\text{and}\quad\langle\Delta w^{A}\Delta
w^{B}\rangle=\eta m^{AB}\Delta t~. \label{fluc}%
\end{equation}

I finish this section with two remarks. The first is on \emph{the nature of
clocks}: In Newtonian mechanics time is defined to simplify the motion of free
particles. The prototype of a clock is a free particle: it moves equal
distances in equal times. In ED time is also defined to simplify the dynamics
of free particles. The prototype of a clock is a free particle too: as we see
in (\ref{fluc}) \emph{free particles} (because for sufficiently short times
all particles are free) \emph{undergo equal fluctuations in equal times}.

The second remark is on \emph{the nature of mass}. As we shall soon see the
constants $m_{n}$ will be identified with the particles' masses. Then,
eq.(\ref{fluc}) provides an interpretation of what `mass' is: mass is an
inverse measure of fluctuations. It is not unusual to treat the concept of
mass as an unexplained primitive concept that measures the amount of stuff (or
perhaps the amount of energy) and then state that quantum fluctuations are
inversely proportional to mass. In an inference scheme such as ED the presence
of fluctuations do not require an explanation. They reflect uncertainty, the
natural consequence of incomplete information. This opens the door to an
entropic explanation of mass: \emph{mass is just a measure of uncertainty
about the expected motion}.

\section{Diffusive dynamics}

The dynamics of $\rho(x,t)$, given by the integral equation (\ref{CK b}), is
more conveniently re-written in a differential form known as the Fokker-Planck
(FP) equation \cite{Reif 1965},
\begin{equation}
\partial_{t}\rho=-\partial_{A}\left(  b^{A}\rho\right)  +\frac{1}{2}\eta
m^{AB}\partial_{A}\partial_{B}\rho~. \label{FP a}%
\end{equation}
(For the algebraic details see \emph{e.g.}, \cite{Caticha 2012}.) The FP
equation can also be written as a continuity equation,
\begin{equation}
\partial_{t}\rho=-\partial_{A}\left(  \rho v^{A}\right)  ~. \label{FP b}%
\end{equation}
The product $\rho v^{A}$ in (\ref{FP b}) represents the probability current,
and $v^{A}$ is interpreted as the velocity of the probability flow --- it is
called the \emph{current velocity}. From (\ref{FP a}) the current velocity in
(\ref{FP b}) is the sum of two separate contributions,
\begin{equation}
v^{A}=b^{A}+u^{A}\,. \label{curr a}%
\end{equation}
The first term is the drift velocity $b^{A}\propto\partial^{A}S$ in
(\ref{drift velocity}) and describes the tendency of the distribution
$\rho(x)$ to evolve so as to increase the entropy of the $y$ variables. One
could adopt a language that resembles a causal mechanism: \emph{it is as if
}the system were pushed by an entropic force. But, of course, such a
mechanistic language should not be taken literally. Strictly speaking,
entropic forces cannot be causal agents because they are purely epistemic.
They might influence our beliefs and expectations but they are not physically
capable of pushing the system --- the \textquotedblleft real
causes,\textquotedblright\ if any, lie elsewhere. Or perhaps, there are no
real causes. In classical mechanics, for example, there is nothing out there
that causes a free particle to persist in its uniform motion in a straight
line. Similarly, in ED we are not required to identify what makes changes
happen the way they do.

The second term\ in (\ref{curr a}) is the \emph{osmotic velocity},
\begin{equation}
u^{A}=-\eta m^{AB}\partial_{B}\log\rho^{1/2}~.
\end{equation}
It represents diffusion, the tendency for probability to flow down the density
gradient. Indeed, one might note that the osmotic or diffusive component of
the current, $\rho u^{A}$, obeys a version of Fick's law in configuration
space,
\begin{equation}
\rho u^{A}=-\frac{1}{2}\eta m^{AB}\partial_{B}\rho=-D^{AB}\partial_{B}\rho
\end{equation}
where $D^{AB}=\eta m^{AB}/2$ is the diffusion tensor \cite{Reif 1965}. Here
too one might be tempted to adopt a causal mechanistic language and say that
the diffusion is driven by the fluctuations expressed in eq.(\ref{fluc}).
\emph{It is as if }the system were bombarded by some underlying random field.
But eq.(\ref{fluc}) need not represent actual fluctuations; it represents our
uncertainty about where the particle will be found after $\Delta t$. The mere
fact that we happen to be uncertain about the position of the particles does
not imply that something must be shaking them.\footnote{E. T. Jaynes warned us
that \textquotedblleft the fact that our information is able to determine [a
quantity] $F$ to 5 percent accuracy, is not enough to make it fluctuate by 5
percent!\textquotedblright\ This mistake he called the Mind Projection Fallacy
\cite{Jaynes 1990}.}

Next we note that since both $b^{A}$ and $u^{A}$ are gradients, their sum ---
the current velocity --- is a gradient too,%
\begin{equation}
v^{A}=m^{AB}\partial_{B}\Phi\quad\text{where}\quad\Phi=\eta(S-\log\rho
^{1/2})~. \label{curr}%
\end{equation}
The FP equation,
\begin{equation}
\partial_{t}\rho=-\partial_{A}\left(  \rho m^{AB}\partial_{B}\Phi\right)  ~,
\label{FP c}%
\end{equation}
can be conveniently rewritten in yet another equivalent but very suggestive
form involving functional derivatives. For some suitably chosen functional
$\tilde{H}[\rho,\Phi]$ we have
\begin{equation}
\partial_{t}\rho(x)=\frac{\delta\tilde{H}}{\delta\Phi(x)}~. \label{Hamilton a}%
\end{equation}
It is easy to check that the appropriate functional $\tilde{H}$ is%
\begin{equation}
\tilde{H}[\rho,\Phi]=\int dx\,\frac{1}{2}\rho m^{AB}\partial_{A}\Phi
\partial_{B}\Phi+F[\rho]~, \label{Hamiltonian a}%
\end{equation}
where the integration constant $F[\rho]$ is some unspecified functional of
$\rho$.

We have just exhibited our first example of a mechanics without mechanism: a
standard diffusion derived from principles of entropic inference. Next we turn
our attention to the derivation of quantum mechanics.

\section{Hamiltonian dynamics}

The previous discussion has led us to a \emph{standard} diffusion. It involves
a single dynamical field, the probability density $\rho(x)$, that evolves in
response to a fixed non-dynamical \textquotedblleft
potential\textquotedblright\ given by the entropy of the $y$ variables,
$S(x)$. In contrast, a \emph{quantum} dynamics consists in the coupled
evolution of two dynamical fields: the density $\rho(x)$ and the phase of the
wave function. This second field can be naturally introduced into ED by
allowing the entropy $S(x)$ to become dynamical:\ the entropy $S$ guides the
evolution of $\rho$, and in return, the evolving $\rho$ reacts back and
induces a change in $S$. This amounts to an entropic dynamics in which each
short step is constrained by a slightly different drift potential; the
constraint (\ref{constraint p}) is continuously updated at each instant in time.

Clearly, different updating rules lead to different forms of ED. The rule that
turns out be particularly useful in physics is inspired by an idea of Nelson's
\cite{Nelson 1979}. We require that $S$ be updated in such a way that a
certain functional, later to be called \textquotedblleft
energy,\textquotedblright\ remains constant. Such a rule may appear to be
natural --- how could it be otherwise? Could we possibly imagine physics
without a conserved energy? But this naturalness is deceptive because ED is
not at all like a classical mechanics. Indeed, the classical interpretation of
a Langevin equation such as (\ref{Delta x}) is that of a Brownian motion in
the limit of infinite friction. This means that in order to provide a
classical explanation of quantum behavior we would need to assume that the
particles were subjected to \emph{infinite} friction while undergoing
\emph{zero} dissipation. Such a strange dynamics could hardly be called
`classical'; the relevant information that is captured by our choice of
constraints cannot be modelled by invoking some underlying classical mechanism
--- this is mechanics without a mechanism. Furthermore, while it is true that
an updating rule based on the notion of a conserved total energy happens to
capture the relevant constraints for a wide variety of physics problems, its
applicability is limited even within physics. For example, in the curved
spacetimes that are used to model gravity it is not possible to even define a
global energy, much less require its global conservation. (For the updating
rules that apply to curved spaces see \cite{Ipek Abedi Caticha 2016}.)

\paragraph{The ensemble Hamiltonian --}

In the standard approach to mechanics one starts with an action and from its
invariance under time translations one derives the conservation of energy. Our
derivation proceeds in the opposite direction: we first identify energy
conservation as the relevant piece of information and from it we derive
Hamilton's equations and their associated action principle.

It turns out that the empirically successful energy functionals are of the
form (\ref{Hamiltonian a}). We impose that, irrespective of the initial
conditions, the entropy $S$ or, equivalently, the potential $\Phi$ in
(\ref{curr}), will be updated in such a way that the functional $\tilde
{H}[\rho,\Phi]$ in (\ref{Hamiltonian a}) is always conserved,
\begin{equation}
\tilde{H}[\rho+\delta\rho,\Phi+\delta\Phi]=\tilde{H}[\rho,\Phi]~,
\end{equation}
or,
\begin{equation}
\frac{d\tilde{H}}{dt}=\int dx\,\left[  \frac{\delta\tilde{H}}{\delta\Phi
}\partial_{t}\Phi+\frac{\delta\tilde{H}}{\delta\rho}\partial_{t}\rho\right]
=0~.
\end{equation}
Using eq.(\ref{Hamilton a}) we get
\begin{equation}
\frac{d\tilde{H}}{dt}=\int dx\,\left[  \partial_{t}\Phi+\frac{\delta\tilde{H}%
}{\delta\rho}\right]  \partial_{t}\rho=0~. \label{dHdt}%
\end{equation}
We want $d\tilde{H}/dt=0$ to hold for arbitrary choices of the initial values
of $\rho$ and $\Phi$. Using eq.(\ref{FP c}) this translates into requiring
$d\tilde{H}/dt=0$ for arbitrary choices of $\partial_{t}\rho$. Therefore,
$\Phi$ must be updated according to
\begin{equation}
\partial_{t}\Phi=-\frac{\delta\tilde{H}}{\delta\rho}~. \label{Hamilton b}%
\end{equation}
Equations (\ref{Hamilton a}) and (\ref{Hamilton b}) are recognized as a
conjugate pair of Hamilton's equations and the conserved functional $\tilde
{H}[\rho,\Phi]$ in (\ref{Hamiltonian a}) is then called the \emph{ensemble
Hamiltonian}.

Thus, the form of the ensemble\ Hamiltonian $\tilde{H}$ is chosen so that the
first Hamilton equation (\ref{Hamilton a}) is the FP eq.(\ref{FP b}), and then
the second Hamilton equation (\ref{Hamilton b}) becomes a generalized
Hamilton-Jacobi equation,
\begin{equation}
\partial_{t}\Phi=-\frac{\delta\tilde{H}}{\delta\rho}=-\frac{1}{2}%
m^{AB}\partial_{A}\Phi\partial_{B}\Phi-\frac{\delta F}{\delta\rho}~.
\label{HJ}%
\end{equation}
This is our second example of a mechanics without mechanism: \emph{a
non-dissipative ED leads to Hamiltonian dynamics}.

\paragraph*{The action --}

We have just seen that the field $\rho(x)$ is a generalized coordinate and
$\Phi(x)$ is its canonical momentum. Now that we have Hamilton's equations,
(\ref{Hamilton a}) and (\ref{Hamilton b}), it is straightforward to invert the
usual procedure and \emph{construct} an action principle from which they can
be derived. Just define the differential
\begin{equation}
\delta A=\int dt\int dx\left[  \left(  \partial_{t}\rho-\,\frac{\delta
\tilde{H}}{\delta\Phi}\right)  \delta\Phi-\left(  \partial_{t}\Phi
+\frac{\delta\tilde{H}}{\delta\rho}\right)  \delta\rho\right]
\end{equation}
and then integrate to get an \textquotedblleft action\textquotedblright,
\begin{equation}
A[\rho,\Phi]=\int dt\left(  \int dx\,\Phi\dot{\rho}-\tilde{H}[\rho
,\Phi]\right)  ~.
\end{equation}
Thus, by construction, imposing $\delta A=0$ leads to (\ref{Hamilton a}) and
(\ref{Hamilton b}). Thus, in the ED approach, actions are not particularly
fundamental; they are just clever ways to summarize the dynamics in a very
condensed form.

\section{Information geometry and the Quantum Potential}

Different choices of the functional $F[\rho]$ in (\ref{Hamiltonian a}) lead to
different dynamics. Earlier we used information geometry, eq.(\ref{gamma C}),
to define the metric $m_{AB}$ of configuration space. Here we use information
geometry once again to motivate the particular choice of the functional
$F[\rho]$ that leads to quantum theory.

The special role played by the particle positions leads us to consider the
family of distributions $\rho(x|\theta)$ that are generated from a
distribution $\rho(x)$ by translations in configuration space by a vector
$\theta^{A}$, $\rho(x|\theta)=\rho(x-\theta)$. The extent to which
$\rho(x|\theta)$ can be distinguished from the slightly displaced
$\rho(x|\theta+d\theta)$ or, equivalently, the information distance between
$\theta^{A}$ and $\theta^{A}+d\theta^{A}$, is given by
\begin{equation}
d\ell^{2}=g_{AB}d\theta^{A}d\theta^{B} \label{dl^2}%
\end{equation}
where%
\begin{equation}
g_{AB}(\theta)=\int dx\frac{1}{\rho(x-\theta)}\frac{\partial\rho(x-\theta
)}{\partial\theta^{A}}\frac{\partial\rho(x-\theta)}{\partial\theta^{B}}~.
\end{equation}
Changing variables $x-\theta\rightarrow x$ yields%
\begin{equation}
g_{AB}(\theta)=\int dx\frac{1}{\rho(x)}\frac{\partial\rho(x)}{\partial x^{A}%
}\frac{\partial\rho(x)}{\partial x^{B}}=I_{AB}[\rho]~. \label{Fisher}%
\end{equation}
Note that these are translations in configuration space. They are not
translations in which the system is displaced as a whole in $3$-d space by the
same constant amount, $(\vec{x}_{1},\vec{x}_{2},\ldots\vec{x}_{N}%
)\rightarrow(\vec{x}_{1}+\vec{\varepsilon},\vec{x}_{2}+\vec{\varepsilon
},\ldots\vec{x}_{N}+\vec{\varepsilon})$. The metric (\ref{Fisher}) measures
the extent to which a distribution $\rho(x)$ can be distinguished from another
distribution $\rho^{\prime}(x)$ in which \emph{just one particle} has been
slightly shifted while all others remain untouched, \emph{e.g.}, $(\vec{x}%
_{1},\vec{x}_{2},\ldots\vec{x}_{N})\rightarrow(\vec{x}_{1}+\vec{\varepsilon
},\vec{x}_{2},\ldots\vec{x}_{N})$.

The simplest choice of functional $F[\rho]$ is linear in $\rho$, $F[\rho]=\int
dx\,\rho(x)V(x)$, and the function $V(x)$ will play the role of the familiar
scalar potential. In an \emph{entropic} dynamics one might also expect
contributions that are of a purely informational nature. Information geometry
provides us with two tensors: one is the metric of configuration space
$\gamma_{AB}\propto m_{AB}$, and the other is $I_{AB}[\rho]$. The simplest
nontrivial scalar that can be constructed from them is the trace $m^{AB}%
I_{AB}$. This suggests
\begin{equation}
F[\rho]=\xi m^{AB}I_{AB}[\rho]+\int dx\,\rho(x)V(x)~,~ \label{QP a}%
\end{equation}
where $\xi>0$ is a constant that controls the relative strength of the two
contributions. The case $\xi<0$ leads to instabilities and is therefore
excluded. (From eq.(\ref{Fisher}) we see that $m^{AB}I_{AB}$ is a contribution
to the energy such that those states that are more smoothly spread out tend to
have lower energy.) The case $\xi=0$ leads to a qualitatively different theory
--- a hybrid dynamics that is both indeterministic and yet classical
\cite{Bartolomeo Caticha 2015}. The term $m^{AB}I_{AB}$ is usually called the
\textquotedblleft quantum\textquotedblright\ potential or the
\textquotedblleft osmotic\textquotedblright\ potential.\footnote{To my
knowledge the relation between the quantum potential and the Fisher
information was first pointed out in \cite{Reginatto 1998}.} It is the crucial
term that accounts for all quintessentially `quantum' effects ---
superposition, entanglement, wave packet expansion, tunnelling, and so on.

With the choice (\ref{QP a}) for $F[\rho]$ the generalized Hamilton-Jacobi
equation (\ref{HJ}) becomes
\begin{equation}
-\partial_{t}\Phi=\frac{1}{2}m^{AB}\partial_{A}\Phi\partial_{B}\Phi+V-4\xi
m^{AB}\frac{\partial_{A}\partial_{B}\rho^{1/2}}{\rho^{1/2}}~. \label{HJb}%
\end{equation}

\section{The Schr\"{o}dinger equation}

Once we have the coupled equations (\ref{FP c}) and (\ref{HJb}) we are done
--- it may not yet be obvious yet but this is quantum mechanics. Purely for
the sake of convenience it is useful to combine $\rho$ and $\Phi$ into a
single complex function
\begin{equation}
\Psi_{k}=\rho^{1/2}\exp(i\Phi/k)\,,~ \label{psi k}%
\end{equation}
where $k$ is an arbitrary positive constant which amounts to rescaling the
constant $\eta$ in eq.(\ref{curr}). Then the two equations (\ref{FP c}) and
(\ref{HJb}) can be written into a single complex equation,
\begin{equation}
ik\partial_{t}\Psi_{k}=-\frac{k^{2}}{2}m^{AB}\partial_{A}\partial_{B}\Psi
_{k}+V\Psi_{k}+\left(  \frac{k^{2}}{2}-4\xi\right)  m^{AB}\frac{\partial
_{A}\partial_{B}|\Psi_{k}|}{|\Psi_{k}|}\Psi_{k}~, \label{sch b}%
\end{equation}
which is quite simple and elegant except for the last non-linear term. It is
at this point that we can take advantage of our freedom in the choice of $k$.
Since the dynamics is fully specified through $\rho$ and $\Phi$ the different
choices of $k$ in $\Psi_{k}$ all lead to different versions of the same
theory. Among all these equivalent descriptions it is clearly to our advantage
to pick the $k$ that is most \emph{convenient} --- a process sometimes known
as `regraduation'.\footnote{Other notable examples of regraduation include the
Kelvin choice of absolute temperature and the Cox derivation of the sum and
product rule for probabilities \cite{Caticha 2012}.} The optimal choice,
$k_{\text{opt}}=(8\xi)^{1/2}$, is such that the non-linear term drops out and
is identified with Planck's constant,
\begin{equation}
\hbar=(8\xi)^{1/2}~.
\end{equation}
Then eq.(\ref{sch b}) becomes the Schr\"{o}dinger equation,%
\begin{equation}
i\hbar\partial_{t}\Psi=-\frac{\hbar^{2}}{2}m^{AB}\partial_{A}\partial_{B}%
\Psi+V\Psi=%
{\displaystyle\sum\limits_{n}}
\frac{-\hbar^{2}}{2m_{n}}\nabla_{n}^{2}\Psi+V\Psi~, \label{sch c}%
\end{equation}
where the wave function is%
\begin{equation}
\Psi=\rho e^{i\Phi/\hbar}~.
\end{equation}
The constant $\xi=\hbar^{2}/8\ $in eq.(\ref{QP a}) turned out to play a
crucial role: it defines the numerical value of what we call Planck's constant
and sets the scale that separates quantum from classical regimes.

The conclusion is that for any positive value of the constant $\xi$ it is
always possible to combine $\rho$ and $\Phi$ to a physically equivalent but
more convenient description where the Schr\"{o}dinger equation is linear. From
the ED perspective the linear superposition principle together with its
attendant complex linear Hilbert spaces are definitely important. But they are
important because they are convenient calculational tools and not because they
are fundamental.

\section{Some final comments}

A theory such as ED can lead to many questions. Here are a few.

\subsubsection*{Is ED equivalent to quantum mechanics?}

Are the Fokker-Planck eq.(\ref{FP c}) and the generalized Hamilton-Jacobi
eq.(\ref{HJb}) fully equivalent to the Schr\"{o}dinger equation? This
question, first raised by Wallstrom \cite{Wallstrom 1989} in the context of
Nelson's stochastic mechanics \cite{Nelson 1985}, concerns the single- or
multi-valuedness of phases and wave functions. Briefly Wallstrom's objection
is that Nelson's stochastic mechanics led to phases $\Phi$ and wave functions
$\Psi$ that are either both multi-valued or both single-valued. Both
alternatives are unsatisfactory: quantum mechanics forbids multi-valued wave
functions, while single-valued phases can exclude physically relevant states
(\emph{e.g.}, states with non-zero angular momentum). We will not discuss this
issue except to note that the objection does not apply once particle spin is
incorporated into ED \cite{Carrara Caticha 2017}\cite{Caticha Carrara 2018}. A
similar argument was developed by Takabayasi in the very different context of
his hydrodynamical approach to quantum theory \cite{Takabayasi 1983}.

\subsubsection*{Is ED a hidden-variable model?}

Let us return to these mysterious auxiliary $y$ variables. Should we think of
them as hidden variables? There is a trivial sense in which the $y$ variables
are \textquotedblleft hidden\textquotedblright: they are not directly
observable.\footnote{The $y$ variables are not observable \emph{at the current
stage of development of the theory}. It may very well happen that once we
learn where to look we will find that they have been staring us in the face
all along.} But being unobserved is not sufficient to qualify as a hidden
variable. The original motivation behind attempts to construct hidden variable
models was to explain or at least ameliorate certain aspects of quantum
mechanics that clash with our classical preconceptions. But the $y$ variables
address none of these problems. Let us mention a few of them:

\textbf{(1) }\emph{Indeterminism}: Is ultimate reality random? Do the gods
play dice? In the standard view quantum theory is considered an extension of
classical mechanics --- indeed, the subject is called quantum \emph{mechanics}
--- and therefore deviations from causality demand an explanation. In the
entropic view, on the other hand, \emph{quantum theory is not mechanics; it is
inference --- }entropic inference is a framework designed to handle
insufficient information. From the entropic perspective indeterminism requires
no explanation. Uncertainty and probabilities are the norm; it is the
certainty and determinism of the classical limit that demand an explanation
\cite{Demme Caticha 2016}.

\textbf{(2) }\emph{Non-classical mechanics}: A common motivation for hidden
variables is that a sub-quantum world will eventually be discovered where
nature obeys essentially classical laws. But in ED there is no underlying
classical dynamics --- both quantum and its classical limit are derived. The
peculiar non-classical effects associated with the wave-particle duality arise
not so much from the $y$ variables themselves but rather from the specific
non-dissipative diffusion which leads to a Schr\"{o}dinger equation. The
important breakthrough here was Nelson's realization that diffusion phenomena
could be much richer than previously expected --- non-dissipative diffusions
can account for wave and interference effects.

\textbf{(3) }\emph{Non-classical probabilities}: It is often argued that
classical probability fails to describe the double slit experiment; this is
not true \cite{Caticha 2012}. It is the whole entropic framework --- and not
the $y$ variables --- that is incompatible with the notion of quantum
probabilities. From the entropic perspective it makes just as little sense to
distinguish quantum from classical probabilities as it is would be to talk
about economic or medical probabilities.

\textbf{(4) }\emph{Non-locality}: Realistic interpretations of the wave
function often lead to such paradoxes as the wave function collapse and the
non-local Einstein-Podolski-Rosen (EPR) correlations \cite{Schlosshauer
2004}\cite{Jaeger 2009}. Since in the ED approach the particles have definite
positions and we have introduced auxiliary $y$ variables that might resemble
hidden variables it is inevitable that one should ask whether this theory
violates Bell inequalities? Or, to phrase the question differently: where
precisely is non-locality introduced? The answer is that the theory has been
formulated directly in $3N$-dimensional configuration space and the
Hamiltonian has been chosen to include the highly non-local quantum potential
(\ref{QP a}). So, yes, the ED model developed here properly describes the
highly non-local effects that lead to EPR correlations and to violations of
Bell inequalities.

\subsubsection*{On interpretation}

We have derived quantum theory as an example of entropic inference. The
problem of interpretation of quantum mechanics is solved because instead of
starting with the mathematical formalism and then seeking an interpretation
that can be consistently attached to it, one starts with a unique
interpretation and then one builds the formalism. This allows a clear
separation between the ontic and the epistemic elements. In ED there is no
risk of confusing which is which. \textquotedblleft Reality\textquotedblright%
\ is represented through the positions of the particles, and our
\textquotedblleft limited information about reality\textquotedblright\ is
represented by probabilities as they are updated to reflect the physically
relevant\ constraints. In ED all other quantities, including the wave
function, are purely epistemic tools. Even energy and momentum and all other
so-called \textquotedblleft observables\textquotedblright\ are epistemic; they
are not properties of the particles but of the wave functions. \cite{Johnson
Caticha 2011}\cite{Vanslette Caticha 2016}

To reiterate a point we made above: since \textquotedblleft
quantum\textquotedblright\ probabilities were never mentioned one might think
that entropic dynamics is a classical theory. But this is misleading: in
ED\ probabilities are neither classical nor quantum; they are tools for
inference. All those non-classical phenomena, such as the non-local effects
that arise in double-slit interference experiments, or the entanglement that
leads to non-local Einstein-Podolski-Rosen correlations, are the natural
result of the linearity that follows from including the quantum potential term
in the ensemble Hamiltonian.

In ED neither action principles nor Hilbert spaces are fundamental. They are
convenient tools designed to summarize the dynamical laws derived from the
deeper principles of entropic inference. The requirement that an energy be
conserved is an important piece of information (that is, a constraint) which
will be fully justified once the extension of entropic dynamics to gravity is developed.

The derivation of laws of physics as examples of inference has led us to
discuss the concept of time. The notion of entropic time was introduced to
keep track of the accumulation of changes. It includes assumptions about the
concept of instant, of simultaneity, of ordering, and of duration. A question
that is bound to be raised is whether and how entropic time is related to the
actual, real, \textquotedblleft physical\textquotedblright\ time. In a similar
vein, to quantify the uncertainties in the motion --- the fluctuations --- to
each particle we associated one constant $m_{n}$. We are naturally led to ask:
How are these constants related to the masses of the particles?

The answers are provided by the dynamics itself: by deriving the
Schr\"{o}dinger equation from which we can obtain its classical limit
(Newton's equation, $F=ma$ \cite{Demme Caticha 2016}) we have shown that the
$t$ that appears in the laws of physics is entropic time and the constants
$m_{n}$ are masses. The argument is very simple:\ it is the Schr\"{o}dinger
equation and its classical limit that are used to design and calibrate our
clocks and our mass-measuring devices. We conclude that by their very design,
\emph{the time measured by clocks is entropic time}, \emph{and what mass
measurements yield are the constants }$m_{n}$. No notion of time that is in
any way more \textquotedblleft real\textquotedblright\ or more
\textquotedblleft physical\textquotedblright\ is needed. Most interestingly,
even though the dynamics is time-reversal invariant, entropic time is not. The
model automatically includes an arrow of time.

Finally, here we have focused on the derivation of examples of dynamics that
are relevant to physics, but the fact that ED is based on inference methods
that are of universal applicability and, in particular, the fact that in the
entropic dynamics framework one deliberately abstains from framing hypothesis
about underlying mechanisms, suggests that it may be possible to adapt these
methods to fields other than physics \cite{Abedi et al 2018}.

\paragraph*{Acknowledgments}

My views on this subject have benefited from discussions with many students
and collaborators including M. Abedi, D. Bartolomeo, C. Cafaro, N. Caticha, A.
Demme, S. DiFranzo, A. Giffin, S. Ipek, D.T. Johnson, K. Knuth, S. Nawaz, P.
Pessoa, M. Reginatto, C. Rodriguez, and K. Vanslette.


\begin{thebibliography}{99}                                                                                               %


\bibitem {Wheeler Zurek 1983}J. A. Wheeler and W. H. Zurek, \emph{Quantum
Theory and Measurement} (Princeton U. Press, Princeton 1983), p. 203 and 210.

\bibitem {Newton 1693}Isaac Newton's third letter to Bentley, February 25,
1693 in \emph{Isaac Newton's papers and letters on Natural Philosophy and
related documents}, ed. by I. B. Cohen (Cambridge, 1958), p. 302.

\bibitem {Stapp 1972}H.P. Stapp, \textquotedblleft The Copenhagen
Interpretation,\textquotedblright\ Am. J. Phys. \textbf{40}, 1098 (1972).

\bibitem {Schlosshauer 2004}M. Schl\"{o}sshauer, \textquotedblleft%
\textquotedblleft Decoherence, the measurement problem, and interpretations of
quantum mechanics,\textquotedblright\ Rev. Mod. Phys. \textbf{76}, 1267 (2004).

\bibitem {Jaeger 2009}G. Jaeger, \emph{Entanglement, Information, and the
Interpretation of Quantum Mechanics} (Springer-Verlag, Berlin Heidelberg 2009).

\bibitem {Leifer 2014}M. S. Leifer, \textquotedblleft Is the quantum state
real? An extended review of $\psi$-ontology theorems,\textquotedblright%
\ Quanta \textbf{3}, 67 (2014); arXiv:1409.1570.

\bibitem {Nelson 1985}E. Nelson, \emph{Quantum Fluctuations }(Princeton UP,
Princeton, 1985).

\bibitem {de la Pena Cetto 1996}L. de la Pe\~{n}a; A.M. Cetto, \emph{The
Quantum Dice, an Introduction to Stochastic Electrodynamics} (Kluwer,
Dordrecht, Holland, 1996).

\bibitem {Smolin 2006}L. Smolin, \textquotedblleft Could quantum mechanics be
an approximation to another theory?\textquotedblright\ arXiv.org/abs/quant-ph/0609109.

\bibitem {tHooft 2002}G. 't Hooft, \textquotedblleft Determinism beneath
Quantum Mechanics,\textquotedblright\ arxiv:quant-ph/0212095;
\textquotedblleft Emergent quantum mechanics and emergent
symmetries,\textquotedblright\ arXiv:hep-th/0707.4568.

\bibitem {Adler 2004}S. Adler, \emph{Quantum Theory as an Emergent
Phenomenon}, (Cambridge UP, Cambridge, 2004).

\bibitem {Grossing 2008}G. Gr\"{o}ssing, \textquotedblleft The Vacuum
Fluctuation Theorem: Exact Schr\"{o}dinger Equation via Nonequilibrium
Thermodynamics,\textquotedblright\ Phys. Lett. A, \textbf{372}, 4556 (2008),
arXiv:0711.4954; G. Gr\"{o}ssing \emph{et al}, \textquotedblleft The Quantum
as an emergent System,\textquotedblright\ J. Phys. Conf. Ser. \textbf{361},
012008 (2012).

\bibitem {EmQm 2015}See \emph{e.g.}, the proceedings of the conference on
\emph{EmQm15: Emergent Quantum Mechanics} \emph{2015}, J. Phy. Conf. Ser.
\textbf{701} (2016) -- http://iopscience.iop.org/issue/1742-6596/701/1.

\bibitem {Caticha 2010}A. Caticha, \textquotedblleft Entropic dynamics, time,
and quantum theory,\textquotedblright\ J. Phys. A: Math. Theor. \textbf{44},
225303 (2011); arXiv.org/abs/1005.2357.

\bibitem {Caticha 2014}A. Caticha, \textquotedblleft Entropic Dynamics: an
inference approach to quantum theory, time and measurement,\textquotedblright%
\ J. Phys.: Conf. Ser. \textbf{504}, 012009 (2014); arXiv:1403.3822.

\bibitem {Caticha et al 2014}A. Caticha, D. Bartolomeo, and M. Reginatto,
\textquotedblleft Entropic Dynamics: from entropy and information geometry to
Hamiltonians and quantum mechanics,\textquotedblright\ in \emph{Bayesian
Inference and Maximum Entropy Methods in Science and Engineering}, ed. by A.
Mohammad-Djafari and F. Barbaresco, AIP Conf. Proc. \textbf{1641}, 155 (2015); arXiv:1412.5629.

\bibitem {Caticha 2015}A. Caticha, \textquotedblleft Entropic
Dynamics\textquotedblright, Entropy \textbf{17}, 6110 (2015); arXiv:1509.03222.

\bibitem {Jaynes 1957}E. T. Jaynes, \textquotedblleft Information Theory and
Statistical Mechanics\textquotedblright, Phys. Rev. \textbf{106}, 620 and
\textquotedblleft Information Theory and Statistical Mechanics
II,\textquotedblright\ \textbf{108}, 171 (1957);

\bibitem {Jaynes 1983}\emph{E. T. Jaynes: Papers on Probability, Statistics
and Statistical Physics}, Ed. by R. D. Rosenkrantz (Reidel, Dordrecht, 1983).

\bibitem {Jaynes 2003}E. T. Jaynes, \emph{Probability Theory: The Logic of
Science} edited by G. L. Bretthorst (Cambridge UP, 2003).

\bibitem {Caticha 2012}A. Caticha, \emph{Entropic Inference and the
Foundations of Physics} (EBEB 2012, S\~{a}o Paulo, Brazil); http://www.albany.edu/physics/ACaticha-EIFP-book.pdf.

\bibitem {Wootters 1981}W. K. Wootters, \textquotedblleft Statistical distance
and Hilbert space,\textquotedblright\ Phys. Rev. D \textbf{23}, 357-362 (1981).

\bibitem {Caticha 1998}A. Caticha, \textquotedblleft Consistency and Linearity
in Quantum Theory,\textquotedblright\ Phys. Lett. A \textbf{244}, 13-17
(1998); Consistency, Amplitudes, and Probabilities in Quantum
Theory\textquotedblright\ ; Phys. Rev. A \textbf{57}, 1572-1582 (1998);
\textquotedblleft Insufficient Reason and Entropy in Quantum
Theory,\textquotedblright\ Found. Phys. \textbf{30}, 227-251 (2000).

\bibitem {Brukner Zeilinger 2002}C. Brukner, A. Zeilinger, \textquotedblleft
Information and Fundamental Elements of the Structure of Quantum
Theory\textquotedblright\ in\ \emph{Time, Quantum, Information}, ed. L.
Castell, and O. Ischebeck, (Springer 2003); arXiv:quant-ph/0212084.

\bibitem {Spekkens 2007}R. Spekkens, \textquotedblleft Evidence for the
epistemic view of quantum states: A toy theory,\textquotedblright\ Phys. Rev.
A \textbf{ 75}, 032110 (2007).

\bibitem {Goyal Knuth Skilling 2010}P. Goyal, K. Knuth, and J. Skilling,
\textquotedblleft Origin of complex quantum amplitudes and Feynman's
rules,\textquotedblright\ Phys. Rev. A \textbf{81}, 022109 (2010).

\bibitem {Hardy 2011}L. Hardy, \textquotedblleft Reformulating and
Reconstructing Quantum Theory,\textquotedblright\ arXiv:1104.2066.

\bibitem {Hall Reginatto 2002}M.J.W. Hall, M. Reginatto, \textquotedblleft
Schr\"{o}dinger equation from an exact uncertainty
principle,\textquotedblright\ J. Phys. A \textbf{35}, 3289-3299 (2002);
\textquotedblleft Quantum mechanics from a Heisenberg-type
inequality,\textquotedblright\ Fortschr. Phys. \textbf{50}, 646-656 (2002).

\bibitem {Reginatto 2013}M. Reginatto, \textquotedblleft From information to
quanta: a derivation of the geometric formulation of quantum theory from
information geometry,\textquotedblright\ arXiv:1312.0429.

\bibitem {Chiribela et al 2011}G. Chiribella, G. M. D'Ariano, and P.
Perinotti, \textquotedblleft Informational derivation of quantum
theory,\textquotedblright\ Phys. Rev. \textbf{A 84}, 012311 (2011).

\bibitem {DAriano 2017}G. M. D'Ariano, \textquotedblleft Physics without
physics: the power of information-theoretical principles,\textquotedblright%
\ Int. J. Th. Phys. \textbf{56}, 97 (2017)

\bibitem {Johnson Caticha 2011}D.T. Johnson and A. Caticha, \textquotedblleft
Entropic dynamics and the quantum measurement problem,\textquotedblright%
\ \ \emph{Bayesian Inference and Maximum Entropy Methods in Science and
Engineering}, ed. by K. Knuth \emph{et al.}, AIP Conf. Proc. \textbf{1443},
104 (2012); arXiv:1108.2550

\bibitem {Vanslette Caticha 2016}K. Vanslette and A. Caticha,
\textquotedblleft Quantum measurement and weak values in entropic quantum
dynamics,\textquotedblright\ in \emph{Bayesian Inference and Maximum Entropy
Methods in Science and Engineering}, ed. by G. Verdoolaege, AIP Conf. Proc.
1853, 090003 (2017); arXiv:1701.00781.

\bibitem {Nawaz Caticha 2011}S. Nawaz and A. Caticha, \textquotedblleft
Momentum and uncertainty relations in the entropic approach to quantum
theory,\textquotedblright\ \ \emph{Bayesian Inference and Maximum Entropy
Methods in Science and Engineering}, ed. by K. Knuth \emph{et al.}, AIP Conf.
Proc. \textbf{1443}, 112 (2012); arXiv:1108.2629.

\bibitem {Bartolomeo Caticha 2015}D. Bartolomeo and A. Caticha,
\textquotedblleft Entropic Dynamics: the Schr\"{o}dinger equation and its
Bohmian limit,\textquotedblright\ \emph{Bayesian Inference and Maximum Entropy
Methods in Science and Engineering}, ed. by A.Giffin and K. Knuth, AIP Conf.
Proc. \textbf{1757}, 030002 (2016); arXiv:1512.09084.

\bibitem {Bartolomeo Caticha 2016}D. Bartolomeo and A. Caticha,
\textquotedblleft Trading drift and fluctuations in entropic dynamics: quantum
dynamics as an emergent universality class,\textquotedblright\ J. Phys: Conf.
Series \textbf{701}, 012009 (2016); arXiv:1603.08469.

\bibitem {Demme Caticha 2016}A. Demme and A. Caticha, \textquotedblleft The
classical limit of entropic quantum dynamics,\textquotedblright\ in
\emph{Bayesian Inference and Maximum Entropy Methods in Science and
Engineering}, ed. by G. Verdoolaege, AIP Conf. Proc. \textbf{1853}, 090001
(2017); arXiv.org:1612.01905.

\bibitem {Nawaz et al  2015}S. Nawaz, M. Abedi, and A. Caticha,
\textquotedblleft Entropic Dynamics in Curved Spaces,\textquotedblright%
\ \emph{Bayesian Inference and Maximum Entropy Methods in Science and
Engineering}, ed. by A.Giffin and K. Knuth, AIP Conf. Proc. \textbf{1757},
030004 (2016); arXiv:1601.01708.

\bibitem {Ipek Caticha 2014}S. Ipek and A. Caticha, \textquotedblleft Entropic
Quantization of Scalar Fields\textquotedblright, \emph{Bayesian Inference and
Maximum Entropy Methods in Science and Engineering}, ed. by A.
Mohammad-Djafari and F. Barbaresco, AIP Conf. Proc. \textbf{1641}, 345 (2015); arXiv:1412.5637.

\bibitem {Ipek Abedi Caticha 2016}S. Ipek, M. Abedi, and A. Caticha,
\textquotedblleft Entropic dynamics of scalar quantum fields: a manifestly
covariant approach,\textquotedblright\ in\ \emph{Bayesian Inference and
Maximum Entropy Methods in Science and Engineering}, ed. by G. Verdoolaege,
AIP Conf. Proc. \textbf{1853}, 090002 (2017).

\bibitem {Caticha 2017}A. Caticha, \textquotedblleft Entropic Dynamics:
Quantum Mechanics from Entropy and Information Geometry,\textquotedblright\ to
appear in Annalen der Physik (2018); arXiv:1711.02538v2. 

\bibitem {Price 1996}H. Price, \emph{Time's Arrow and Archimedes' Point}
(Oxford UP, New York 1996).

\bibitem {Zeh 2002}H. D. Zeh, \emph{The Physical Basis of the Direction of
Time} (Springer-Verlag, 2002).

\bibitem {Amari 1985}S. Amari, \emph{Differential-Geometrical Methods in
Statistics} (Springer-Verlag, 1985).

\bibitem {Lanczos 1970}C. Lanczos, \emph{The Variational Principles of
Mechanics} (Dover, New York 1986).

\bibitem {Reif 1965}F. Reif, \emph{Fundamentals of Statistical and Thermal
Physics} (McGraw-Hill, 1965).

\bibitem {Jaynes 1990}E. T. Jaynes, \textquotedblleft Probability in Quantum
Theory\textquotedblright, in\ \emph{Complexity, Entropy and the Physics of
Information}, ed. by W. H Zurek (Addison-Welsey, Reading MA, 1990) and online
at http://bayes.wustl.edu.

\bibitem {Nelson 1979}E. Nelson, \textquotedblleft Connection between Brownian
motion and quantum mechanics,\textquotedblright\ \emph{Einstein Symposium
Berlin}, Lect. Notes Phys. \textbf{100}, p.168 (Springer-Verlag, Berlin, 1979).

\bibitem {Reginatto 1998}M. Reginatto, \textquotedblleft Derivation of the
equations of nonrelativistic quantum mechanics using the principle of minimum
Fisher information,\textquotedblright\ Phys. Rev. \textbf{A 58}, 1775 (1998).

\bibitem {Wallstrom 1989}T. C. Wallstrom, \textquotedblleft On the derivation
of the Schr\"{o}dinger equation from stochastic mechanics,\textquotedblright%
\ Found. Phys. Lett. \textbf{2}, 113 (1989); \textquotedblleft The
inequivalence between the Schr\"{o}dinger equation and the Madelung
hydrodynamic equations,\textquotedblright\ Phys. Rev. \textbf{A 49}, 1613 (1994).

\bibitem {Carrara Caticha 2017}N. Carrara and A. Caticha, \textquotedblleft
Quantum phases in entropic dynamics,\textquotedblright\ presented at MaxEnt
2017, 37th International Workshop on Bayesian Inference and Maximum Entropy
Methods in Science and Engineering (July 9-14, 2017, Jarinu, Brazil); arXiv:1708.08977.

\bibitem {Caticha Carrara 2018}A. Caticha and N. Carrara, \textquotedblleft
The entropic dynamics of spin,\textquotedblright\ in preparation.

\bibitem {Takabayasi 1983}T. Takabayasi, \textquotedblleft Vortex, spin, and
triad for quantum mechanics of spinning particle,\textquotedblright\ Prog.
Theor. Phys. \textbf{70}, 1-17 (1983).

\bibitem {Abedi et al 2018}M. Abedi, D. Bartolomeo, and A. Caticha,
\textquotedblleft Modeling the stock market as an entropic
dynamics,\textquotedblright\ in preparation.
\end{thebibliography}
\end{document}